\documentclass[superscriptaddress,aps,twocolumn]{revtex4}
\usepackage{amssymb}
\usepackage{amsmath}
\usepackage{graphicx}
\usepackage{color}
\usepackage{amsfonts}
\usepackage{multirow}

\begin{document}

\title{Decoherent phonon effects in fast atom-surface scattering}
\author{L. Frisco}
\affiliation{Instituto de Astronom\'{\i}a y F\'{\i}sica del Espacio (UBA-CONICET), Ciudad
Universitaria, (C1428EGA) Buenos Aires, Argentina.}
\author{M.S. Gravielle\thanks{%
Author to whom correspondence should be addressed.\newline
Electronic address: msilvia@iafe.uba.ar}}
\affiliation{Instituto de Astronom\'{\i}a y F\'{\i}sica del Espacio (UBA-CONICET), Ciudad
Universitaria, (C1428EGA) Buenos Aires, Argentina.}
\date{\today }

\begin{abstract}
The study of the influence of phonon-mediated processes on grazing-incidence
fast atom diffraction (GIFAD) patterns is relevant for the use of GIFAD as a
surface analysis technique. In this work, we apply the Phonon-Surface
Initial Value Representation (P-SIVR) approximation to investigate lattice
vibration effects on GIFAD patterns for the He/LiF(001) system at room
temperature. The main features introduced by thermal lattice vibrations in
the angular distributions of the scattered helium atoms are analyzed by
considering normal energies in the 0.1 -3 eV range. In all the cases,
thermal fluctuations introduce a wide polar spread that transforms the
interference maxima into elongated strips. We found that the polar width of
these maxima does not depend on the normal energy, as it was experimentally
observed. In addition, when the normal energy increases, not only the
relative intensities of interference peaks are affected by the crystal
vibrations, but also the visibility of the interference structures, which
disappear completely for normal energies approximately equal to 3 eV. These
findings qualitatively agree with recent experimental data, but the
simulated polar widths underestimate the experimentally-derived limit,
suggesting that there are other mechanisms, such as inelastic phonon
processes, that contribute to the polar dispersion of the GIFAD patterns.
\end{abstract}

\maketitle



\section{Introduction}

Grazing-incidence fast atom diffraction (GIFAD) has been the subject of
extensive experimental and theoretical research since its first experimental
reports in 2007 \cite{Schuller2007,Rousseau2007}. From the theoretical point
of view, significant advances have been reached in the description of GIFAD
over these 15 years \cite{Diaz2022}. Different theoretical methods, which
range from classical and semiclassical (or semiquantum) approaches to full
quantum calculations, have been applied to describe and analyze GIFAD
experiments for a wide variety of crystal targets, not only at room
temperature but also at higher temperatures \cite{Atkinson2014}. But in the
vast majority of these works, diffraction patterns were assumed to be
produced by elastic scattering from ideal frozen crystal surfaces, with the
atoms at rest at their equilibrium positions, whereas the contribution of
lattice vibrations, i.e., phonons, was considered to play a minor role, as
suggested in early articles \cite{Aigner2008,Rousseau2008,Schuller2010}. \
However, the influence of phonon-mediated processes on GIFAD patterns is
still not fully understood, becoming a topic that has recently attracted
renewed interest \cite{Roncin2017,Schram2018,
Frisco2019,Frisco2020,Debiossac2021,Pan2021,Pan2022}.

The most remarkable effect introduced\ by phonons in GIFAD patterns from
insulator surfaces is a wide polar dispersion of the projectile distribution
\cite{Frisco2019,Pan2021}, a feature that was early predicted by Manson
\textit{et al. }\cite{Manson2008}. While atomic projectiles elastically
scattered off an ideal static surface hit the detector plane forming a thin
annulus, named the Laue circle, as a consequence of the energy conservation
\cite{Schuller2007,Rousseau2007}, the inclusion of lattice vibrations within
the theoretical models transforms the interference maxima into elongated
streaks along the polar angle, as it is usually observed in GIFAD
experiments \cite{Frisco2019,Frisco2020,Debiossac2021,Pan2021}. Hence, the
polar\ profile of GIFAD patterns was the focus of recent experimental and
theoretical works, where the variation of the polar-profile width as a
function of both, the normal incidence energy \cite{Pan2021} and the
sample's temperature \cite{Frisco2020,Pan2022}, was investigated. On the
other hand, a systematic theoretical study of thermal vibration effects on
the azimuthal characteristics of the interference patterns as a function of
the incidence conditions is still missing.

In this paper, we use the He/LiF(001) system at room temperature, which is a
prototype of the GIFAD phenomenon, as a benchmark to study how the phonon
effects associated with the thermal fluctuations of the LiF lattice vary as
the incidence conditions change. The aim is to analyze the dependence of
such thermal effects on the normal incidence energy $E_{\bot }$ $=E\sin
^{2}\theta _{i}$, with \ $E$ being the total incidence energy and $\theta
_{i}$ the glancing incidence angle. Notice that in typical GIFAD
experiments, the interference structures of the projectile distributions are
mainly governed by\ $E_{\bot }$ \cite{Schuller2009,Muzas2016}. Consequently,
the relative intensities of the interference maxima \ as a function of $%
E_{\bot } $ are commonly employed to determine the electronic or
morphological characteristics of the crystal surface \cite{Winter2011}. This
fact makes it relevant to know the influence of the normal energy in the
thermal vibration effects on the diffraction patterns.

Our study is based on the Phonon-Surface Initial Value Representation
(P-SIVR) approximation \cite{Frisco2019}, which is a semiquantum method that
accounts for phonon transitions using a quantum harmonic crystal model.
Within the P-SIVR approach, the scattering probability is expressed as a
series in terms of the net number of phonons exchanged during the
atom-surface collision. The first-order term, P0-SIVR, associated with the
zero-phonon scattering process \cite{Frisco2019,Frisco2020}, is here applied
to investigate the $E_{\bot }$- dependence of the lattice vibration effects
on the angular distributions of He atoms grazingly scattered off a LiF(001)
surface along the $\left\langle 110\right\rangle $ channel. The incidence
conditions are chosen in accord with a recent experimental study for
He/LiF(001) GIFAD by Debiossac \textit{et al.} \cite{Debiossac2021}, which
will be used for comparison.

The article is organized as follows. The P0-SIVR approach is summarized in
Sec. II, while results are presented and discussed in Sec. III, analyzing
the decoherent limit and the phonon effects on the polar profile. In Sec. IV
we outline our conclusions. Atomic units (a.u.) are used unless otherwise
stated.

\section{Theoretical model}

The P-SIVR approximation is based on the previous Surface Initial Value
Representation (SIVR) approach for grazing scattering from a rigid surface
\cite{Gravielle2014}, which has been successfully applied to different
GIFAD\ problems \cite{Bocan2016,Gravielle2015,Frisco2018,Bocan2020,Bocan2021}%
. The basic idea of \ the\ P-SIVR and SIVR methods is to use the Van Vleck
representation of the time-evolution quantum operator, as given by Miller
\cite{Miller2001}, within the frame of a time-dependent distorted wave
formalism \cite{Dewangan1994}. This strategy makes it possible to describe
the quantum interference effects involved in GIFAD in terms of classical
trajectories with different initial conditions, incorporating an
approximated representation of the classically forbidden transitions that
contribute to the dark side of the rainbow angle. But in contrast with the
SIVR approach, where the surface is represented by means of an ideal static
crystal model, the P-SIVR approximation makes use of the harmonic crystal
description \cite{Ashcroft}, which allows us to take into account
phonon-mediated processes.

Within the P-SIVR approach, the differential scattering probabilities
associated with transitions between different initial \ and final crystal
states are averaged over the equilibrium distribution of the initial phonon
states, after adding over all the possible final states. Thus, the resulting
P-SIVR probability can be expanded as a series on the net number $n$ of
phonons emitted or absorbed during the collision, where the term for $n=0$,
P0-SIVR, corresponds to the elastic scattering without net phonon exchange,
but including intermediated phonon transitions.

The P0-SIVR probability for the scattering $\mathbf{K}_{i}\longrightarrow $ $%
\mathbf{K}_{f}$, with $\mathbf{K}_{i}$ ( $\mathbf{K}_{f}$) being the initial
\ (final) projectile momentum, can be expressed as a function of an
effective transition amplitude $\mathcal{A}^{(P0-SIVR)}$ as \cite{Frisco2019}
\begin{equation}
\frac{dP^{(P0-SIVR)}}{d\Omega _{f}}=K_{f}^{2}\left\vert \mathcal{A}%
^{(P0-SIVR)}\right\vert ^{2},  \label{Pdif}
\end{equation}%
where $K_{f}=K_{i}$ as a result of the energy conservation and $\Omega
_{f}=(\theta _{f},\varphi _{f})$ determines the $\mathbf{K}_{f}$ direction,
with $\theta _{f}$ and $\varphi _{f}$ being\ the final polar and azimuthal
angles measured with respect to the surface plane and the axial channel,
respectively. The effective P0-SIVR amplitude reads \cite{Frisco2019}
\begin{eqnarray}
\mathcal{A}^{(P0-SIVR)} &=&\int d\mathbf{R}_{o}\ f(\mathbf{R}_{o})\int d%
\mathbf{K}_{o}\ g(\mathbf{K}_{o})  \notag \\
&&\times \ \ \int d\underline{\mathbf{u}}_{o}\ a_{0}(\mathbf{R}_{o},\mathbf{K%
}_{o},\underline{\mathbf{u}}_{o}),  \label{An0}
\end{eqnarray}%
where the functions $f$ $\ $and $g$ describe the position and momentum
profiles, respectively, of the\ incident projectile wave-packet, and $a_{0}(%
\mathbf{R}_{o},\mathbf{K}_{o},\underline{\mathbf{u}}_{o})$ is the partial
amplitude corresponding to the classical projectile trajectory $\mathbf{R}%
_{t}\equiv \mathbf{R}_{t}(\mathbf{R}_{o},\mathbf{K}_{o},\underline{\mathbf{u}%
}_{o})$, which starts at the initial time $t=0$ in the position $\mathbf{R}%
_{o}$ with momentum $\mathbf{K}_{o}$. The trajectory $\mathbf{R}_{t}$
depends also on the spatial configuration $\underline{\mathbf{u}}_{o}$ of
the crystal at $t=0$, where the underlined vector $\underline{\mathbf{u}}%
_{o} $ denotes the $3N$-dimension vector associated with the spatial
displacements of the $N$ ions contained in the crystal sample, with respect
to their equilibrium positions. Such crystal deviations are considered
invariable during the collision time, which is much shorter than the
characteristic time of phonon vibrations \cite{Ashcroft}. In addition, the
partial amplitude $a_{0}(\mathbf{R}_{o},\mathbf{K}_{o},\underline{\mathbf{u}}%
_{o})$ includes an explicit dependence on $\underline{\mathbf{u}}_{o}$
through the momentum-dependent Debye-Waller factor, which acts as an
effective screening of the atom-surface interaction.~The complete expression
of this partial amplitude, along with the steps used to derive the P0-SIVR
approach, can be found in Ref. \cite{Frisco2019}.

\section{Results}

Our study is confined to $5$ keV $^{4}$He atoms grazingly impinging on a
LiF(001) surface along the $\left\langle 110\right\rangle $ channel, with
different normal energies in the $0.1$-$3$ eV range. For these collision
system and incidence conditions experimental angular distributions were
recently reported in Fig. 16 of Ref. \cite{Debiossac2021}.

To determine the influence of phonon effects, in all the cases the final
projectile distributions obtained from the P0-SIVR approach, which includes
intermediate phonon transitions, are compared with those derived from the
SIVR approximation, which assumes an ideal rigid crystal surface. Both
calculations - P0-SIVR and SIVR - were carried out using the same binary
interatomic potentials, taken from Ref. \cite{Miraglia2017}, to build the
pairwise additive surface potential. Furthermore, since the general features
of the GIFAD patterns depend on the collimating conditions of the incident
beam \cite{Seifert2015}, we chosen\ a fixed collimation scheme, formed by a
square slit of size $d$ $=0.09$ mm placed at a distance $L=36$ cm from the
surface, with an angular beam dispersion of $0.006$ deg. These collimating
parameters, which are in accord with current experimental setups for GIFAD
\cite{Bocan2020,Debiossac2021}, were used to evaluate the spatial and
momentum profiles of the incident wave-packet as given in Ref. \cite%
{Gravielle2015}. Details of the P0-SIVR and SIVR calculations can be
respectively found in Refs. \cite{Frisco2019,Frisco2020} and \cite%
{Gravielle2015,Gravielle2018}.

\subsection{Dependence of the phonon effects on $E_{\bot }$}

\begin{figure}[tbp]
\includegraphics[width=0.5 \textwidth]{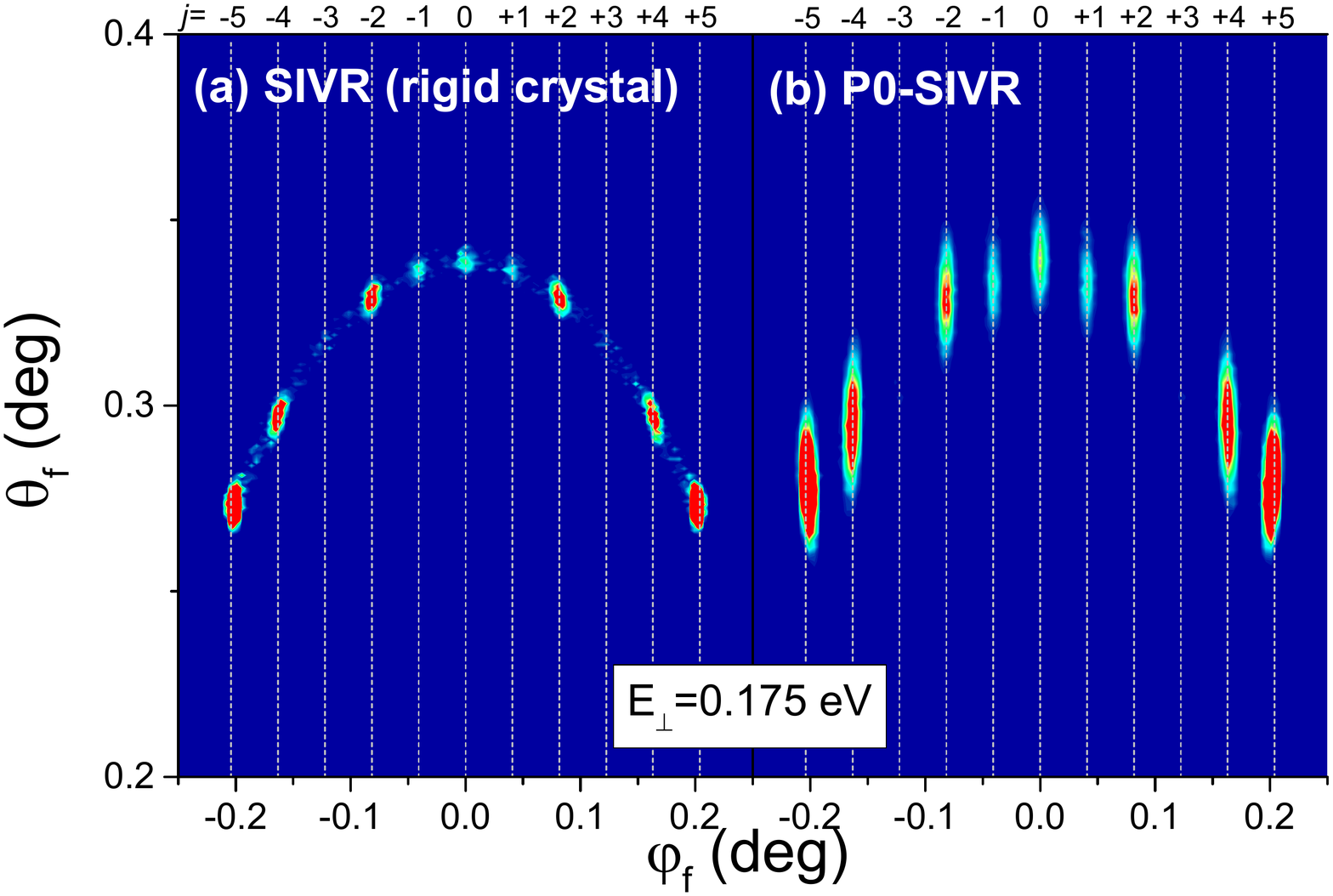} \centering
\caption{(Color online) Two-dimensional projectile distributions, as a
function of $\protect\theta _{f}$ and $\protect\varphi _{f}$, for $5$ keV $%
^{4}$He atoms scattered off LiF(001) at the temperature $T=300$ K. Incidence
along the $\left\langle 110\right\rangle $ channel with the normal energy $%
E_{\perp }=0.175$ eV is considered. Results derived within (a) the SIVR
approximation, for a rigid crystal, and (b) the P0-SIVR approach, including
thermal vibrations, are displayed. Vertical dashed lines indicate the ideal
positions of the Bragg peaks, with orders $j=0,\pm 1,..$.}
\label{map-ep0175}
\end{figure}

We start by considering a low normal energy for He/LiF GIFAD \cite%
{Schuller2009}, $E_{\bot }$ $=0.175$ eV, for which helium projectiles probe
regions far from the topmost crystal layer, with He-surface distances $%
Z\succeq 1.9$ {\AA }, suggesting a minor role of lattice vibrations. In
Figs. \ref{map-ep0175} (a) and \ref{map-ep0175} (b) we respectively show the
SIVR and P0-SIVR two-dimensional distributions for $E_{\bot }$ $=0.175$ eV,
as a function of\ the final polar and azimuthal angles. Except for the
different polar spreads, both angular distributions are very alike: They
present similar interference structures, with less intense Bragg peaks of
order $j=0,\pm 1$,\ and missing $j=\pm 3$ peaks. Moreover, the azimuthal
positions of the Bragg maxima ($\varphi _{j}^{(B)}$, with $j=0,\pm 1,\pm
2,.. $) are not affected by the thermal fluctuations of the crystal atoms,
being $\varphi _{j}^{(B)}=\arcsin (j\lambda /a_{y})$, with $\lambda =2\pi
/K_{i}$ and $a_{y}=5.4$ a.u. the channel width. This latter behavior was
found to be independent of the $E_{\bot }$ value, being related to the
average of the spatial deviations of a large number of crystal atoms
involved in the scattering process. However, despite the low normal energy
considered in Fig. \ref{map-ep0175}, the lattice vibrations produce a wide
polar dispersion of the Bragg maxima, transforming the small spots on the
Laue circle, displayed in Fig. \ref{map-ep0175} (a), into vertical streaks,
as it is observed in Fig. \ref{map-ep0175} (b).

\begin{figure*}[tbp]
\includegraphics[width=0.8 \textwidth]{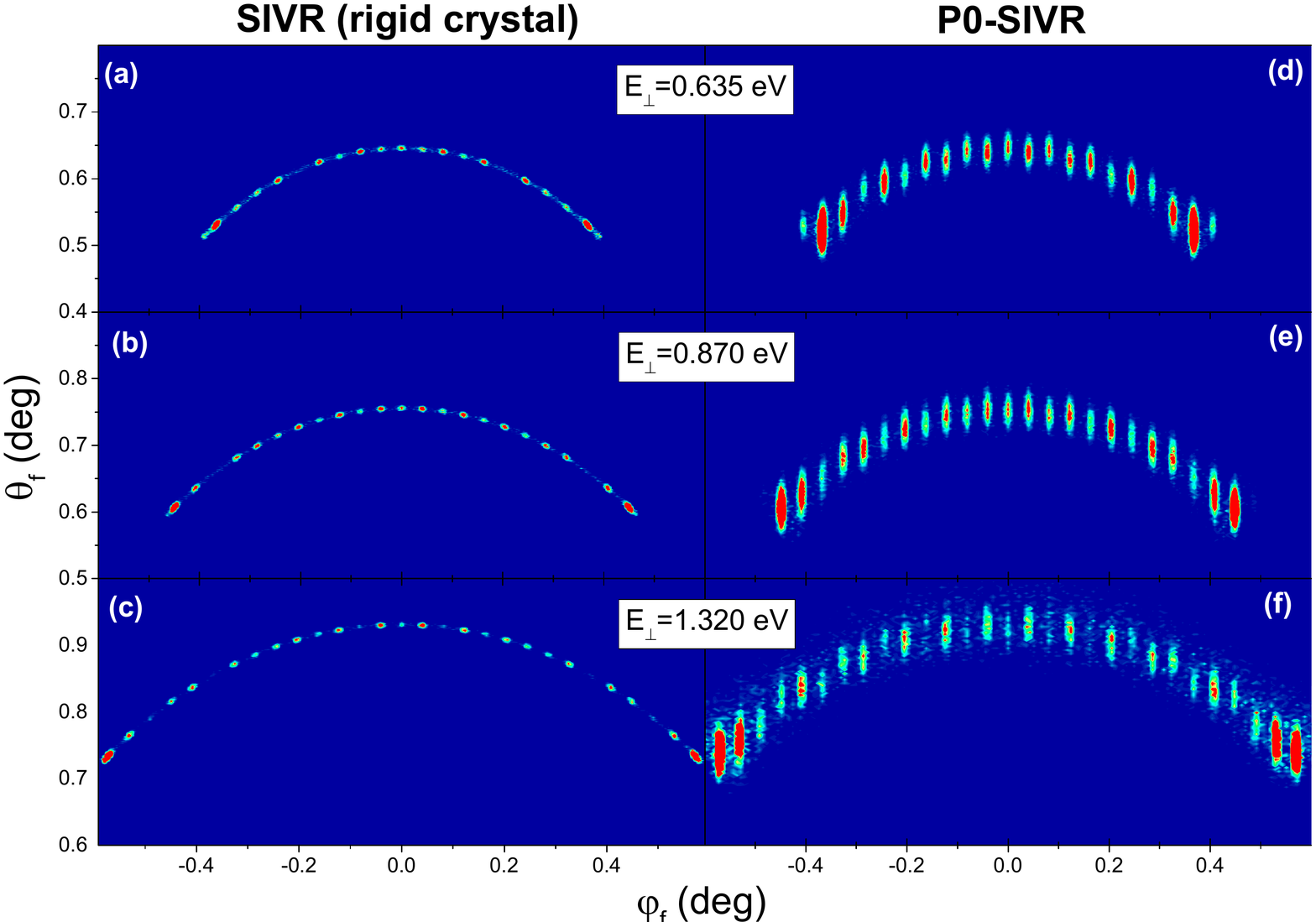} \centering
\caption{(Color online) Analogous to Fig. \protect\ref{map-ep0175} for the
following normal energies: (a) and (d), $E_{\perp }=0.635$ eV; (b) and (e), $%
E_{\perp }=0.870$ eV; and (c) and (f), $E_{\perp }=1.320$ eV. Panels (a),
(b), and (c) (left column) show angular distributions for a rigid crystal
derived within the SIVR approximation, while panels (d), (e), and (f) (right
column) display P0-SIVR angular distributions including effects due to
thermal lattice vibrations.}
\label{map-ep}
\end{figure*}

In Fig. \ref{map-ep} we compare SIVR and P0-SIVR distributions for higher
normal energies: $E_{\bot }$ $=0.635$, $0.870$, and $1.320$ eV. When $%
E_{\bot }$ increases, more equally $\varphi _{f}$-spaced Bragg peaks are
visible in the angular distributions. But again, the most noticeable effect
introduced by lattice vibrations is the polar dispersion of the GIFAD
patterns. The P0-SIVR distributions for the different $E_{\bot }$ values
(right-column \ of Fig. \ref{map-ep}) display a broad polar spread that
gives rise to a thick annulus, with mean radius \ $\theta _{i}$, which
contrast with the thin Laue circle observed for the SIVR distributions
derived by assuming a rigid crystal (left-column). \ Furthermore, the
P0-SIVR interference structures start to blur as \ $E_{\bot }$ becomes
higher than $1$ eV, as a consequence of the decoherent effect produced by
phonon contributions. For $E_{\bot }$ $=1.320$ eV, the P0-SIVR distribution
of Fig. \ref{map-ep} (f) shows a more diffuse GIFAD pattern than those for\
the lower $E_{\bot }$ values, shown in the upper panels of Fig. \ref{map-ep}%
. Instead, the SIVR distributions exhibit well defined sharp maxima in all
the panels of Fig. \ref{map-ep}, without smudging for high normal energies.

\subsection{Decoherent phonon effects}

\begin{figure}[tbp]
\includegraphics[width=0.5\textwidth]{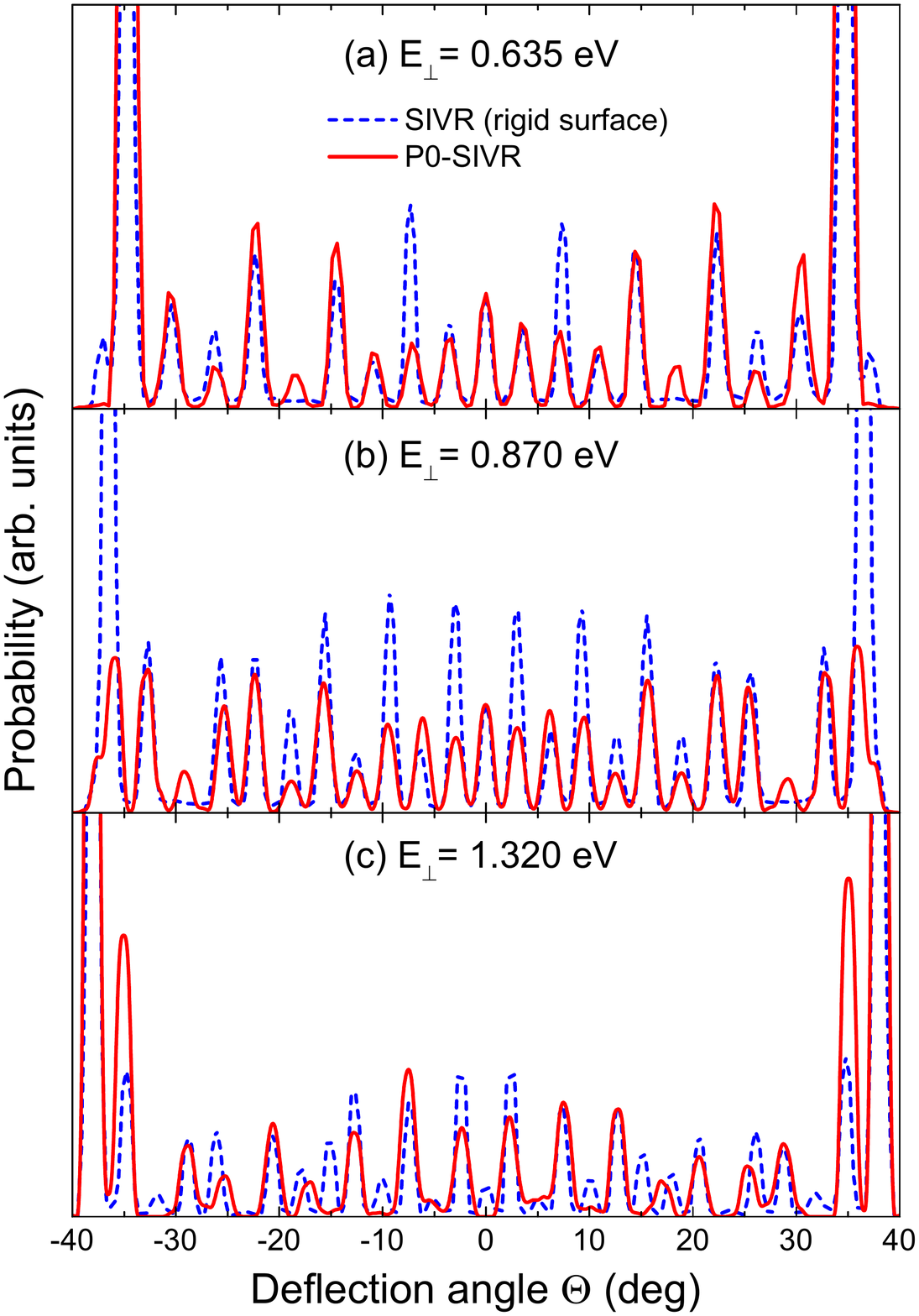} \centering
\caption{(Color online) Differential probabilities as a function of the
deflection angle $\Theta =\arctan (\protect\varphi _{f}/\protect\theta _{f})$%
, for different normal energies: (a) $E_{\perp }=0.635$ eV, (b) $E_{\perp
}=0.870$ eV, and (c) $E_{\perp }=1.320$ eV. In all the panels, red solid
line, P0-SIVR probability including thermal vibrations; blue dashed line,
SIVR probability for a rigid crystal.}
\label{spectra-ep}
\end{figure}

In order to analyze in more detail the decoherent effect of lattice
vibrations on GIFAD patterns, in Fig. \ref{spectra-ep} we plot the projected
intensities, as a function of the deflection angle $\Theta =\arctan (\varphi
_{f}/\theta _{f})$, for the cases of Fig. \ref{map-ep}. For $E_{\bot }$ $%
=0.635$, $0.870$, and $1.320$ eV, the differential P0-SIVR probabilities $%
dP^{{\small (P0-SIVR)}}/d\Theta $, derived from Eq. (\ref{Pdif}) by
integrating over a reduced annular area of central thickness
$0.005\deg $ \cite{Debiossac2016}, are displayed along with the
corresponding SIVR values normalized at $\Theta =0$. For the three
normal energies, thermal fluctuations affect the relative
intensities of the interference maxima, mostly decreasing the
intensity of some Bragg peaks. This thermal effect on \ $dP^{{\small
(P0-SIVR)}}/d\Theta $ is partially related to the polar dispersion
of the P0-SIVR maxima because some of them are shifted above or
below the ideal Laue circle, as also observed in the experiment
\cite{Debiossac2021}, affecting the result of the integral over
$\theta _{f}$
involved in the projection method \cite{Debiossac2016}. Moreover, for $%
E_{\bot }$ $=$ $1.320$ eV phonon-mediated processes included in the P0-SIVR
approach give rise to a decoherent effect in the $\Theta $- spectrum, which
smudges several interference maxima, as it is seen in Fig. \ref{spectra-ep}
(c).

\begin{figure}[tbp]
\includegraphics[width=0.5 \textwidth]{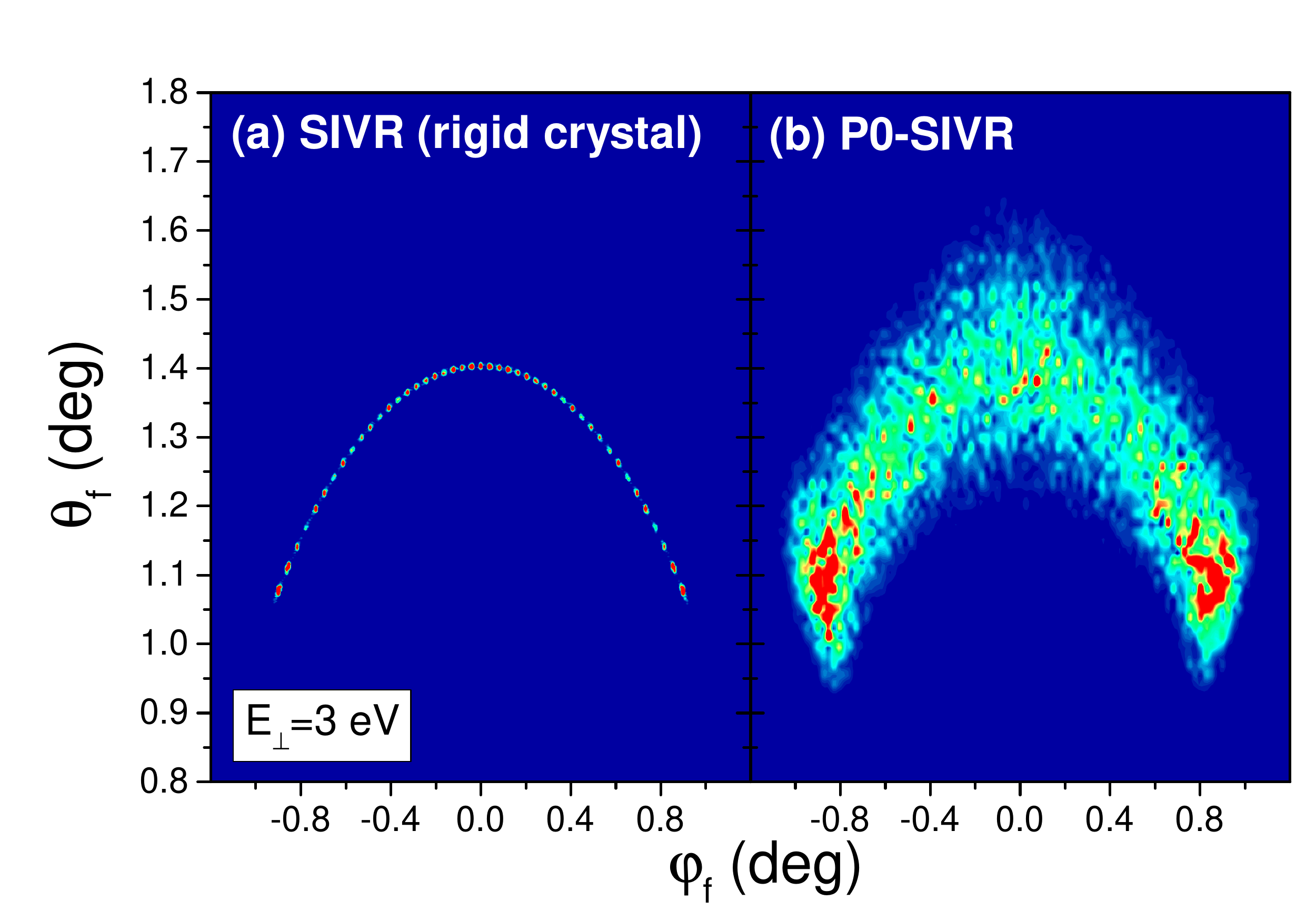} \centering
\caption{(Color online) Analogous to Fig. \protect\ref{map-ep0175} for the
normal energy $E_{\perp }=3$ eV.}
\label{map-ep3}
\end{figure}

Finally, in Fig. \ref{map-ep3} we contrast the angular distributions derived
within the SIVR and P0-SIVR approximations for $E_{\bot }=3$ eV. \ At this
high normal energy, for which helium atoms reach closest approach distances
to the surface of about $Z\simeq 0.85$ {\AA }, the SIVR distribution for a
rigid crystal still exhibits well-defined Bragg maxima over a thin ring
(Fig. \ref{map-ep3} (a)). But remarkably, these interference structures
completely disappear when phonon contributions are taken into account
through the P0-SIVR approach, as it is showed in Fig. \ref{map-ep3} (b).
Notice that at $E_{\bot }=3$ eV, a similar decoherent effect was
experimentally reported in Fig. 16 of Ref. \cite{Debiossac2021}. \
Therefore, these findings suggest that the decoherence introduced by thermal
vibrations when helium projectiles move close to the surface plane is able
to destroy the quantum interference, making the projectile distributions
tend to the classical limit.

In relation to the comparison with the available experimental data, the
P0-SIVR distributions displayed in Figs. \ref{map-ep0175}, \ref{map-ep}, and %
\ref{map-ep3} \ reproduce fairly well the overall features of the
experimental ones shown in Fig. 16 of Ref. \cite{Debiossac2021}. However, we
found visible differences in the relative intensities of the diffraction
maxima for several normal energies. Taking into account that GIFAD patterns
are extraordinarily sensitive to the atom-surface potential, these
discrepancies could be attributed to deficiencies of the pairwise additive
potential model \cite{Miraglia2017}. In particular, at the normal energy $%
E_{\bot }=3$ eV, the absence of any signature of a central peak in the
P0-SIVR distribution of Fig. \ref{map-ep3} (b), in opposition to the
experiment, would suggest that our\ potential model \cite{Miraglia2017} is
not able to properly describe the He-LiF(001) interaction at short
distances. Furthermore, despite the large number of trajectories (more than $%
10^{7}$) used in the P0-SIVR calculations displayed in Fig. \ref{map-ep3}
(b), the rainbow maxima are more diffuse than those observed in the
experimental angular distribution \cite{Debiossac2021}, which shows
outermost oval spots.

\subsection{Phonon effects on the polar profile}

Since the polar dispersion is the main phonon effect on GIFAD patterns, in
this subsection we thoroughly analyze the dependence on the normal energy of
the polar profiles corresponding to different Bragg maxima. In Fig. \ref%
{profile-fi0}\ we plot the polar distribution of the central maximum,
obtained from the P0-SIVR approach, for $E_{\bot }$ $=0.365$, $0.635$, and $%
0.870$ eV. In each panel, the differential probability $dP^{(P0-SIVR)}/d%
\theta _{f}$ at $\varphi _{f}=0$ is compared with the \ SIVR polar profile
derived from a rigid crystal model, which displays a sharp peak at the
specular reflection angle $\theta _{f}=\theta _{i}$. \ In contrast, the
polar distributions derived within the P0-SIVR approach present a broad
asymmetric maximum, peaked at the Laue circle, which is due to the
contribution of phonon-mediated processes. We found that the $\theta _{f}$-
asymmetry with respect to the Laue circle depends on the normal energy, with
a more extended polar distribution below the Laue circle for $E_{\bot }$ $=$
$0.870$ eV, and the inverse behavior for $E_{\bot }$ $=0.365$ and $0.635$
eV.
\begin{figure}[tbp]
\includegraphics[width=0.5\textwidth]{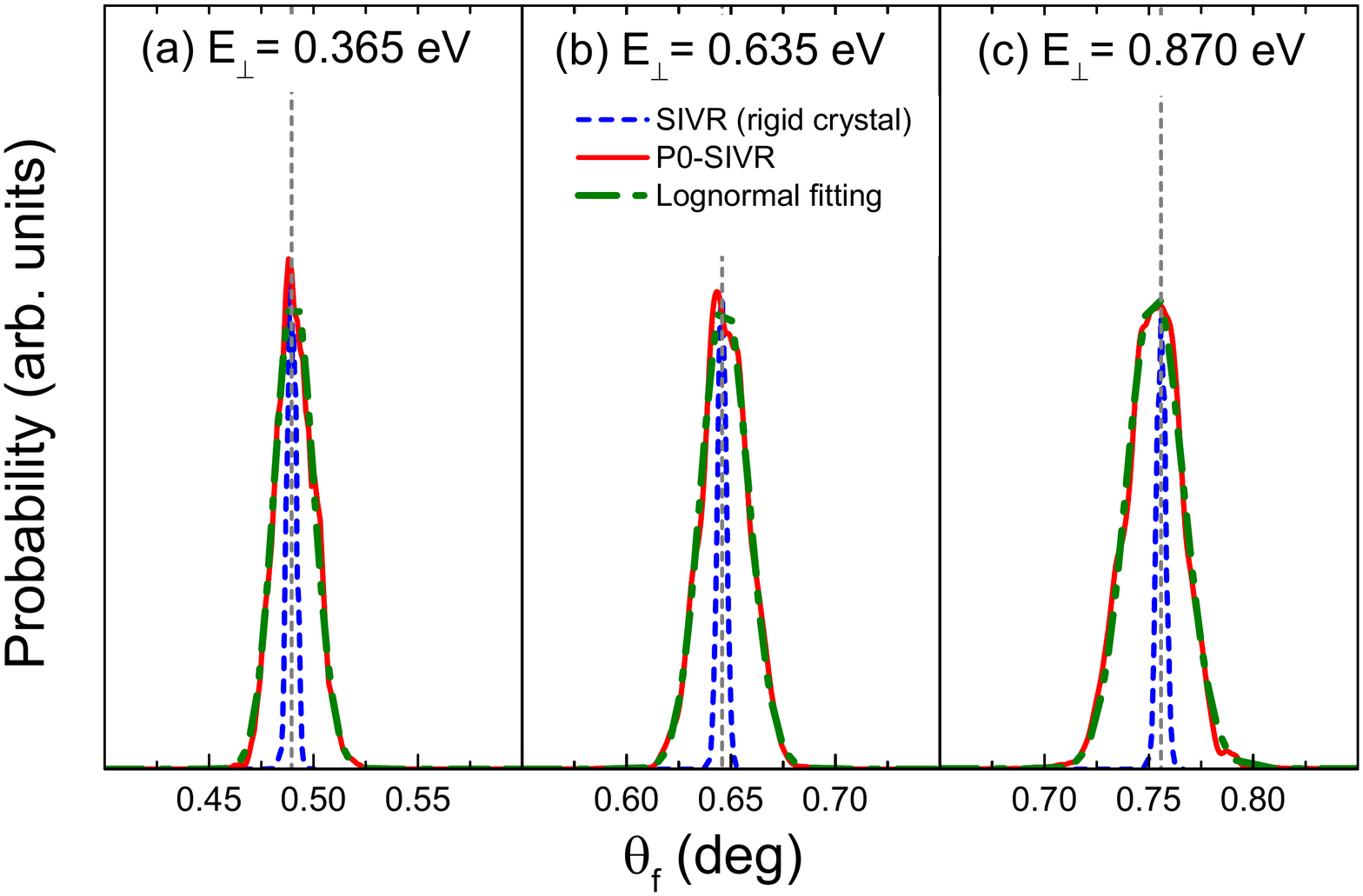} \centering
\caption{(Color online) Intensity profile of the central maximum at $\protect%
\varphi _{f}=0$, as a function of the polar angle $\protect\theta _{f}$, for
different normal energies: (a) $E_{\perp }=0.365$ eV, (b) $E_{\perp }=0.635$
eV, and (c) $E_{\perp }=0.870$ eV. In all the panels, red solid line,
differential probability derived within the P0-SIVR approach; blue dashed
line, SIVR probability for a rigid crystal; green dot-dashed line, lognormal
fitting of the P0-SIVR results, as given by Eq.(\protect\ref{lognormal}).
Vertical gray dashed line, ideal $\protect\theta _{f}$- position on the Laue
circle (i.e., $\protect\theta _{f}=\protect\theta _{i}$).}
\label{profile-fi0}
\end{figure}

\begin{figure}[tbp]
\includegraphics[width=0.5\textwidth]{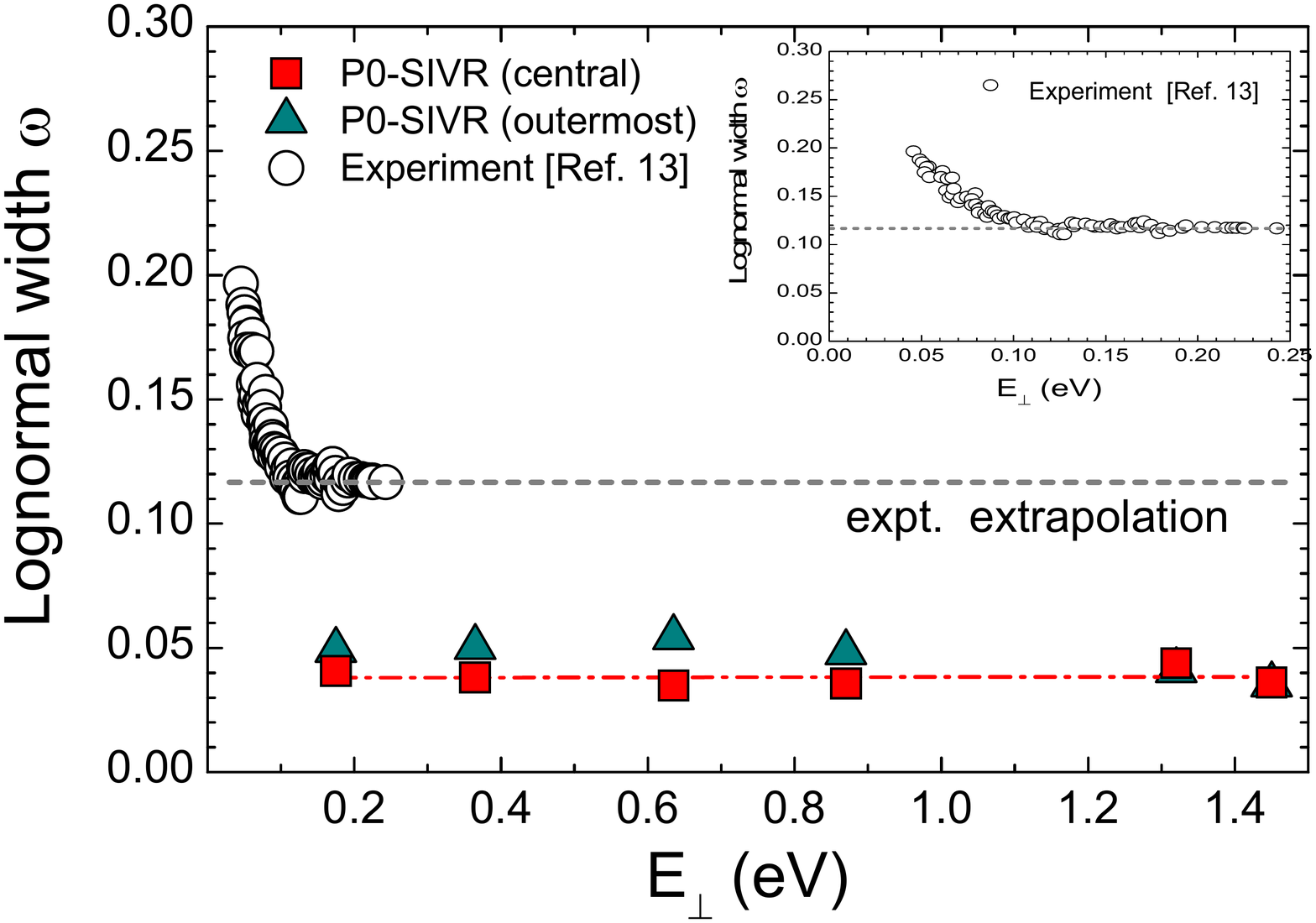} \centering
\caption{(Color online) Lognormal width $\protect\omega$ of the distribution
given by Eq. (\protect\ref{lognormal}), as a function of the normal energy.
Theory: Solid squares and triangles, values derived by fitting P0-SIVR
results for the central and outermost peaks, respectively. The red
dot-dashed line corresponds to the linear fitting of the P0-SIVR lognormal
width of the central peak. Experiment: Empty circles, experimentally-derived
data extracted from Fig. 7 of Ref. \protect\cite{Pan2021}. The black dashed
line indicates the experimental asymptotic limit. Inset: Details of the
experimentally-derived data \protect\cite{Pan2021}.}
\label{ancho-polar}
\end{figure}

In addition, in Fig. \ref{profile-fi0} we also show that the P0-SIVR polar
profiles can be well reproduced by the lognormal function%
\begin{equation}
\mathcal{P}(\theta _{f})=\frac{A}{\omega \ \theta _{f}}\exp [\frac{%
-2(ln(\theta _{f}/\theta _{c}))^{2}}{\omega ^{2}}],  \label{lognormal}
\end{equation}%
as it is usually done to deal with the experimental data \cite%
{Roncin2017,Debiossac2021,Pan2021}, where $A$, $\theta _{c}$, and $\omega $
are fitting parameters. From this fitting procedure we can determine the
lognormal width $\omega $ of the P0-SIVR central peak, which is plotted as a
function of $E_{\bot }$ in Fig. \ref{ancho-polar} (square symbols).
Noticeably, the $\omega $ values derived from the P0-SIVR approach do not
suffer significative changes as the normal energy varies, running along a
horizontal dot-dashed line in\ Fig. \ref{ancho-polar}. A similar polar
behavior was reported in a recent experimental study for a lower normal
energy range \cite{Pan2021}.

As observed in Fig. \ref{ancho-polar}, the experimentally-derived lognormal
widths extracted from Fig. 7 of Ref. \cite{Pan2021}, which are also shown in
the inset \cite{notew}, tend to an asymptotic constant limit as $E_{\bot }$
increases. Then, to compare with these experimental data, obtained under
different incidence conditions, we extrapolate the experimental\ limit to
higher normal energies, finding that the lognormal width of the central peak
of the P0-SIVR distribution underestimates this\ extrapolated limit by a
factor $3$. This fact would indicate the importance of phonon-mediated
scattering with net phonon exchange, which might contribute to the polar
dispersion of the interference patterns, as considered in recent articles
\cite{Roncin2017,Pan2021}. However, notice that the experimentally-derived
data of Ref. \cite{Pan2021} take into account the addition of polar profiles
corresponding to different Bragg orders. If the polar width of the outermost
P0-SIVR peaks is evaluated as a function of $E_{\bot }$, the results,
plotted with triangle symbols in Fig. \ref{ancho-polar}, run a factor 2
below the experimental asymptotic limit, reducing the simulation-experiment
gap.

\section{Conclusions}

In this work, we have study the dependence on the normal energy of the
lattice vibration effects that affect the GIFAD patterns for He/LiF(001) at
room temperature. Our analysis was based on the P0-SIVR approximation \cite%
{Frisco2019}, which is a semiquantum method that describes the zero-phonon
scattering process, including the contribution of intermediate phonon
transitions. In accord with previous experimental and theoretical studies\
\cite{Roncin2017,Frisco2020,Debiossac2021,Pan2021,Pan2022}, we found that
the main effect introduced by the phonon contributions is a wide polar
dispersion of the interference structures, which is present even at low
normal energies. For normal energies in the $0.1$-$1.5$ eV range, the
azimuthal positions of the Bragg peaks coincide with those derived from an
ideal rigid crystal, but the relative intensities of some Bragg orders are
modified by the lattice fluctuations as the normal energy increases,
indicating that thermal effects should be taken into account in order to use
GIFAD as a surface analysis technique.

Furthermore, we observe that the interference patterns start to blur for $%
E_{\bot }$ $\succsim 1.3$ eV, vanishing completely at $E_{\bot }\succsim 3$
eV due to the decoherent effect produced by the thermal vibrations of the
LiF crystal. At high normal energies, analogous non-coherent experimental
distributions were recently reported in Ref. \cite{Debiossac2021}. We found
that the P0-SIVR approach reproduces fairly well the overall features of the
experimental data of Fig. 16 of Ref. \cite{Debiossac2021}. However, it
should be mentioned that in several cases, the relative Bragg intensities of
the simulated patterns are not in full agreement with the experiment. Such
discrepancies might be attributed to the atom-surface interaction, whose
proper description represents a real challenge for the potential models.

Concerning the polar profiles of the P0-SIVR distributions, they are well
fitted with a lognormal function, as it happens with the experiments \cite%
{Pan2021}. Within the P0-SIVR approach, the polar lognormal width $\omega $
of the central maximum remains constant over the $0.1$-$1.5$ eV normal
energy range. This behavior is in agreement with recent
experimentally-derived data for lower normal energies \cite{Pan2021}.
Nevertheless, present P0-SIVR $\omega $- values underestimate the
experimental asymptotic limit by a factor about $2$, which would indicate
that there are other mechanisms, like inelastic phonon-mediated processes
involving net phonon exchange \cite{Debiossac2021,Pan2021}, which might
contribute to the polar spread.

\begin{acknowledgments}
The authors acknowledge financial support from CONICET and ANPCyT of
Argentina.
\end{acknowledgments}

\bibliographystyle{unsrt}
\bibliography{HeLiF-eperp}

\begin{thebibliography}{10}

\bibitem{Schuller2007}
A.~Sch{\"u}ller, S.~Wethekam, and H.~Winter.
\newblock {\em Phys. Rev. Lett.}, 98:016103, 2007.

\bibitem{Rousseau2007}
P.~Rousseau, H.~Khemliche, A.~G. Borisov, and P.~Roncin.
\newblock {\em Phys. Rev. Lett.}, 98:016104, 2007.

\bibitem{Diaz2022}
C.~D\'{\i}az and M.~S. Gravielle.
\newblock {\em Phys. Chem. Chem. Phys.}, 24:15628--15656, 2022.

\bibitem{Atkinson2014}
P.~Atkinson, M.~Eddrief, V.~H. Etgens, H.~Khemliche, M.~Debiossac, A.~Momeni,
  M.~Mulier, B.~Lalmi, and P.~Roncin.
\newblock {\em Appl. Phys. Lett.}, 105:021602, 2014.

\bibitem{Aigner2008}
F.~Aigner, N.~Simonovi\'{c}, B.~Solleder, L.~Wirtz, and J.~Burgd{\"o}rfer.
\newblock {\em Phys. Rev. Lett.}, 101:253201, 2008.

\bibitem{Rousseau2008}
P.~Rousseau, H.~Khemliche, N.~Bundaleski, P.~Soulisse, A.~Momeni, and
  P.~Roncin.
\newblock {\em J. Phys.: Conf. Ser.}, 133:012013, 2008.

\bibitem{Schuller2010}
A.~Sch{\"u}ller, S.~Wethekam, D.~Blauth, H.~Winter, F.~Aigner,
  N.~Simonovi\'{c}, B.~Solleder, J.~Burgd{\"o}rfer, and L.~Wirtz.
\newblock {\em Phys. Rev. A}, 82:062902, 2010.

\bibitem{Roncin2017}
P.~Roncin and M.~Debiossac.
\newblock {\em Phys. Rev. B}, 96:035415, 2017.

\bibitem{Schram2018}
M.~C. Schram and E.~J. Heller.
\newblock {\em Phys. Rev. A}, 98:022137, 2018.

\bibitem{Frisco2019}
L.~Frisco and M.~S. Gravielle.
\newblock {\em Phys. Rev. A}, 100:062703, 2019.

\bibitem{Frisco2020}
L.~Frisco and M.~S. Gravielle.
\newblock {\em Phys. Rev. A}, 102:062821, 2020.

\bibitem{Debiossac2021}
M.~Debiossac, P.~Pan, and P.~Roncin.
\newblock {\em Phys. Chem. Chem. Phys.}, 23:7615--7636, 2021.

\bibitem{Pan2021}
P.~Pan, M.~Debiossac, and P.~Roncin.
\newblock {\em Phys. Rev. B}, 104:165415, 2021.

\bibitem{Pan2022}
P.~Pan, M.~Debiossac, and P.~Roncin.
\newblock {\em Phys. Chem. Chem. Phys.}, 24:12319--12328, 2022.

\bibitem{Manson2008}
J.~R. Manson, H.~Khemliche, and P.~Roncin.
\newblock {\em Phys. Rev. B}, 78:155408, 2008.

\bibitem{Schuller2009}
A.~Sch{\"u}ller, H.~Winter, M.~S. Gravielle, J.~M. Pruneda, and J.~E. Miraglia.
\newblock {\em Phys. Rev. A}, 80:062903, 2009.

\bibitem{Muzas2016}
A.~S. Muzas, F.~Gatti, F.~Mart\'{\i}n, and C.~D\'{\i}az.
\newblock {\em Nucl. Instr. Meth. B}, 382:49--53, 2016.

\bibitem{Winter2011}
H.~Winter and A.~Sch{\"u}ller.
\newblock {\em Prog. Surf. Sci.}, 86:169--221, 2011.

\bibitem{Gravielle2014}
M.~S. Gravielle and J.~E. Miraglia.
\newblock {\em Phys. Rev. A}, 90:052718, 2014.

\bibitem{Bocan2016}
G.~A. Bocan, J.~D. Fuhr, and M.~S. Gravielle.
\newblock {\em Phys. Rev. A}, 94:022711, 2016.

\bibitem{Gravielle2015}
M.~S. Gravielle and J.~E. Miraglia.
\newblock {\em Phys. Rev. A}, 92:062709, 2015.

\bibitem{Frisco2018}
L.~Frisco, J.~E. Miraglia, and M.~S. Gravielle.
\newblock {\em J. Phys.: Condens. Matter}, 30:405001, 2018.

\bibitem{Bocan2020}
G.~A. Bocan, H.~Breiss, S.~Szilasi, A.~Momeni, E.~M. Staicu-Casagrande, M.~S.
  Gravielle, E.~A. S\'{a}nchez, and H.~Khemliche.
\newblock {\em Phys. Rev. Lett.}, 125:096101, 2020.

\bibitem{Bocan2021}
G.~A. Bocan, H.~Breiss, S.~Szilasi, A.~Momeni, E.~M. Staicu-Casagrande, E.~A.
  S\'{a}nchez, M.~S. Gravielle, and H.~Khemliche.
\newblock {\em Phys. Rev. B}, 104:235401, 2021.

\bibitem{Miller2001}
W.~H. Miller.
\newblock {\em J. Phys. Chem A}, 105:2942--2955, 2001, and references therein.

\bibitem{Dewangan1994}
D.~P. Dewangan and J.~Eichler.
\newblock {\em Phys. Rep.}, 247:59--219, 1994.

\bibitem{Ashcroft}
N.~W. Ashcroft and N.~D. Mermin.
\newblock {\em Solid State Physics}.
\newblock Brooks/Cole, Cengage Learning, Belmont, USA, 1976.

\bibitem{Miraglia2017}
J.~E. Miraglia and M.~S. Gravielle.
\newblock {\em Phys. Rev. A}, 95:022710, 2017.

\bibitem{Seifert2015}
J.~Seifert, J.~Lienemann, A.~Sch{\"u}ller, and H.~Winter.
\newblock {\em Nucl. Instrum. Methods Phys. Res. B}, pages 99--105.

\bibitem{Gravielle2018}
M.~S. Gravielle, J.~E. Miraglia, and L.~Frisco.
\newblock {\em Atoms}, 6:64, 2018.

\bibitem{Debiossac2016}
M.~Debiossac and P.~Roncin.
\newblock {\em Nucl. Instr. Methods Phys. Res. B}, 382:36--41, 2016.

\bibitem{notew}
The lognormal widths extracted from Fig. 7 of Ref. 13 were multiplied by 2 in
  accord with Eq. (3).

\end{thebibliography}

\end{document}